\documentclass[11pt,a4paper,onecolumn,reqno]{amsart}
\usepackage[a4paper, margin=2.3cm, bmargin=3cm]{geometry}

\usepackage{graphicx,stfloats} \usepackage{color}
\usepackage{amsmath}
\usepackage{amsaddr}
\usepackage{bm, soul}
\usepackage{mathtools}
\usepackage[colorinlistoftodos,prependcaption,textsize=tiny]{todonotes}
\usepackage{braket}
\usepackage[square,comma,sort&compress, numbers]{natbib}
\usepackage{enumerate}

\usepackage{tikz}
\usepackage{pgfplots}
\pgfplotsset{compat=1.9}
\usetikzlibrary{calc,arrows,fadings,decorations.pathreplacing,decorations.markings,patterns,shapes.geometric}
\tikzset{>=stealth,inner sep=0pt, outer sep=2pt,}
\tikzset{vecteur/.style={->,thick,color=black,smooth}}

\renewcommand{\st}[1]{}

\usepackage{latexsym} \usepackage{subfigure} \usepackage{float}
\usepackage{multirow} 
\usepackage{amssymb}
\usepackage{amsthm}
\usepackage[version=3]{mhchem} \usepackage{siunitx}
 \usepackage{multirow}
\renewcommand\figurename{Fig.~}
\renewcommand\tablename{Table~}

\usepackage{etoolbox}
\patchcmd{\pprintMaketitle}
  {\hrule\vskip12pt}
 {\hrule\vskip12pt\ifvoid\extrainfobox\else\unvbox\extrainfobox\par\vskip12pt\fi}
 {}{}

\newsavebox\extrainfobox

\usepackage{caption}
\renewcommand{\tablename}{Tab.}
\renewcommand{\figurename}{Fig.}
\usepackage{booktabs}
\newcolumntype{K}[1]{>{\centering\arraybackslash}p{#1}}
\newcolumntype{L}[1]{>{\raggedleft\arraybackslash}p{#1}}
\newcolumntype{R}[1]{>{\raggedright\arraybackslash}p{#1}}

\definecolor{orange}{rgb}{1,.65,0}
\definecolor{lightgray}{gray}{0.9}
\definecolor{blue}{HTML}{395FE2}

\newcommand*\tV{\tilde{V}}

\newcommand*\mbf[1]{\mathbf{#1}}

\newcommand{\dd}[2]{\displaystyle\frac{\partial #1}{\partial #2}}
\newcommand{\ddtwo}[2]{\displaystyle\frac{\partial^2 #1}{\partial {#2}^2}}

\title{Can flamelet manifolds capture the interactions of thermo-diffusive instabilities and turbulence in lean hydrogen flames? - An a-priori analysis}

\author{Hannes Böttler$^{a,1,*}$, Driss Kaddar$^{a,1}$, T. Jeremy P. Karpowski$^{a}$, Federica Ferraro$^{a}$, Arne Scholtissek$^{a}$, Hendrik Nicolai$^{a}$, Christian Hasse$^{a}$}
\thanks{$^1$Joint First Authors}
\email{boettler@stfs.tu-darmstadt.de} 
\address{$^a$Technical University of Darmstadt, Department of Mechanical Engineering, Simulation of reactive Thermo-Fluid Systems, Otto-Berndt-Str. 2, 64287 Darmstadt, Germany}

\begin{document}
\pagestyle{plain}

\begin{abstract} 
Flamelet-based methods are extensively used in modeling turbulent hydrocarbon flames. However, these models have yet to be established for (lean) premixed hydrogen flames.
While flamelet models exist for \textit{laminar} thermo-diffusively unstable hydrogen flames, for which consideration of curvature effects has resulted in improved model predictions~\cite{boettler_2022b},
it is still unclear whether these models are directly applicable to turbulent hydrogen flames.
Therefore, a detailed assessment of stretch effects on thermochemical states in a turbulent lean premixed hydrogen-air slot flame through finite-rate chemistry simulations is conducted.
Strain and curvature are examined individually using a composition space model, revealing their distinct influences on thermochemical states. 
An \textit{a-priori} analysis confirms that the previously developed tabulated manifolds fall short of capturing all turbulent flame phenomena,
necessitating a novel manifold incorporating both strain and curvature variations.
These results underscore the significance of these variations in developing manifold-based combustion models for turbulent lean hydrogen flames.
\end{abstract}


\maketitle

\section{Introduction} \addvspace{10pt}
Hydrogen is an energy carrier of high technological relevance. Based on policies and funding programs, a variety of research directions have emerged that focus on hydrogen's economic potential, production, transport, storage, and utilisation~\cite{andersson_2019,dawood_2020,kovac_2021, wappler_2022, kosehoeinghaus_2021, erdener_2023}.  
When generated with renewable energy sources via electrolysis, the so-called green hydrogen can be utilized directly as a fuel or further processed in chemical synthesis.
One scenario is using hydrogen as a fuel in aero-engines and stationary gas turbines, however, its direct usage as a fuel presents additional challenges. Compared to conventional carbon-based fuels, hydrogen exhibits a higher diffusivity and reactivity, leading to significant changes in the combustion dynamics.
Here, a lean (partially) premixed combustion mode would be desirable for low pollutant emissions but can lead to safety issues such as flashback and thermoacoustic oscillations~\cite{lopez_2024,fruzza_2023,zhang_2019}.
To ensure the safe operation of new hydrogen-operated devices, CFD-aided design processes are desirable. To this end, predictive models for turbulent premixed flames are required that take into account the distinct characteristics of hydrogen.

One particular challenge for model development is that lean hydrogen-air flames are prone to intrinsic thermo-diffusive instabilities. Thermo-diffusive instabilities arise due to the significant difference in mass and thermal diffusive fluxes, leading to pronounced differential diffusion effects along the flame front. The ratio of the mass and thermal diffusivity is the Lewis number, which is particularly low for hydrogen $(\mathrm{Le}_\mathrm{H2} \approx 0.3)$~\cite{law_book_2006}.
The strong differential diffusion of hydrogen causes minor perturbations at the flame front to be amplified, resulting in highly wrinkled flame fronts which are characterized by a significantly increased flame speed and substantial fluctuations in local reaction rates. Extensive research in laminar reacting flows has explored these instabilities, revealing that the increase in the overall consumption rate can be attributed to both an enlarged flame surface area as well as an increased fuel consumption rate per unit of flame surface area~\cite{altantzis_2015,frouzakis_2015,berger_2019,creta_2020,wen_2021a, attili_2021, howarth_2022, zhang_2023}.

Considering that industrial-relevant configurations generally feature turbulent flows, the question of how the turbulence-induced flame wrinkling interacts with the thermo-diffusive unstable hydrogen mixture is highly relevant. A set of recently performed DNS studies at various Karlovitz numbers suggests that thermo-diffusive instabilities persist in turbulent flames and are even amplified through synergistic effects~\cite{berger_2022,rieth_2022,rieth_2023}.
Hence, in turbulent hydrogen flames, both turbulence and thermo-diffusive instabilities impact flame wrinkling and local reactivity, ultimately influencing fuel consumption rates and, consequently, turbulent flame speed.
The enhancing effects of turbulence on thermo-diffusive effects are linked to higher strain and flame curvature due to turbulent wrinkling and steepened scalar gradients. Thereby, the differential diffusion of hydrogen is promoted resulting in strong mixture inhomogeneities across the flame front and super-adiabatic temperatures in the exhaust gas~\cite{berger_2022, coulon_2023}. The effects of intrinsic instabilities are further enhanced with increasing pressure~\cite{yang_2018, chu_2023, rieth_2022, howarth_2023}. At high pressure and high Karlovitz numbers, diffusion of molecular and atomic hydrogen still exhibit a leading order effect on the burning rates~\cite{rieth_2022}.
The strong effects of molecular diffusion in turbulent hydrogen flames differ significantly from hydrocarbon fuels, where turbulence-induced flame wrinkling solely determines the turbulent flame speed.

Several numerical studies of turbulent lean premixed \ce{H2}-air slot flames have been based on direct numerical simulations (DNS)~\cite{berger_2022, coulon_2023}.
It has been outlined that these simulations require considerable computational resources. Therefore, reliable models need to be developed to investigate more complex configurations (e.g. technical combustors) at moderate computational cost.
Commonly used modeling approaches are flamelet-based models that incorporate detailed kinetics using tabulated chemistry~\cite{gicquel_2000,pierce_2004,oijen_2016}.
The detailed chemistry information of these flamelets is usually precomputed and stored as a tabulated manifold, which is parameterized by certain control variables. The control variables are related to different physical phenomena, e.g. mixing, reaction progress or non-adiabatic effects. Subsequently, the manifold is coupled to the CFD simulation, where only the transport equations for the control variables are solved and the full thermochemical state is retrieved from the manifold. 

Flamelet-based modeling approaches are well established for turbulent hydrocarbon flames and have been extended to problems of increasingly physical complexity~\cite{fiorina_2015, oijen_2016, popp_2021, nicolai_2022b, steinhausen_2023, kircher_2023, tang_2023, vance_2023, almutairi_2023}.
Recently, several model extensions have been proposed to describe laminar hydrogen-air flames capturing also differential diffusion effects and the subsequent mixture stratification~\cite{schlup_2019, mukundakumar_2021, boettler_2022a, boettler_2022b, lou_2022}.
It is noted that most of these models are usually based on the tabulation of one-dimensional unstretched flamelets and have been developed further with different targets.
Particularly, focusing on capturing flame structures of laminar thermo-diffusively unstable hydrogen-air flames, Böttler et al.~\cite{boettler_2022b} reported that incorporation of curvature effects into the manifold substantially improved the results. In this work, a recently developed composition space model (CSM)~\cite{scholtissek_2019a, scholtissek_2019b} proved beneficial for computing one-dimensional flamelets with curvature variations.
While flamelet-based models have been successfully applied to turbulent methane-hydrogen-air flames~\cite{vreman_2009} and first attempts have been made to adapt turbulence closure models to hydrogen combustion~\cite{luo_2014, lapenna_2021a}, modeling studies on turbulent premixed hydrogen-air flames including thermo-diffusive instabilities are still scarce but highly relevant.
This gap is addressed in this work.

The objective of this work is twofold: (1) A fully resolved simulation (FRS)~\cite{domingo_2023} of a turbulent lean premixed hydrogen-air slot flame is performed using detailed chemistry. Based on this data, the physical phenomena that need to be captured by flamelet-based models to predict the physics of turbulent lean premixed hydrogen-air flames that exhibit thermo-diffusive instabilities are outlined. In particular, the effects of strain and curvature on the thermochemical state are studied individually. (2) Four different flamelet manifolds with increasing complexity are generated, including a novel manifold taking into account both strain and curvature variations. The performance of these manifolds is assessed based on the FRS data by means of an \textit{a-priori} analysis.
The paper is structured as follows: first, the setup of the fully resolved simulation of the lean premixed turbulent hydrogen-air flame is described, followed by an analysis of strain and curvature effects, which are discussed concerning implications for model development. Further, an optimal estimator assessment is performed to elaborate suitable manifold parameterizations.  Thereafter, differently complex flamelet-based modeling approaches are introduced. Finally, this variety of flamelet manifolds is tested against the FRS data in an \textit{a-priori} manner. The paper ends with a conclusion.

\section{Numerical setup} \addvspace{10pt}
\begin{figure}[ht]
    \centering
    \includegraphics[width=65mm]{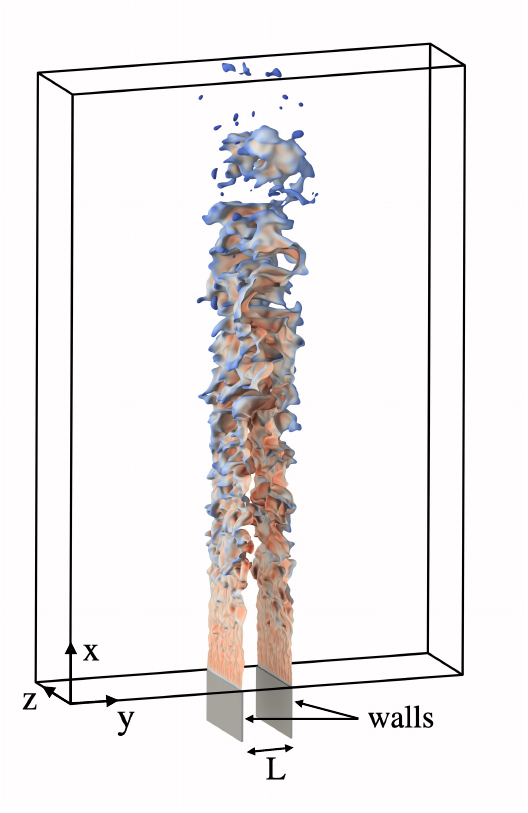}
    \caption{Physical domain of the simulation and visualization of the flame. The flame is represented by a temperature isosurface of $T=\SI{1100}{K}$ and colored by the heat release rate.}
    \label{fig:3D_DNS}
    \end{figure}

\subsection{Configuration and Operating Conditions}

In this work, a turbulent premixed lean hydrogen-air flame in a slot burner configuration is investigated. The unburnt mixture enters the domain through a central jet, which is separated by two walls from the hot coflows. The configuration is inspired by the works of Sankaran et al.~\cite{sankaran_2007} and Berger et al.~\cite{berger_2022}. Figure~\ref{fig:3D_DNS} shows a visualization of the computational domain.
It spans across $12\,L$ in streamwise (x) and crosswise (y) direction and $2\,L$ in spanwise ($z$) direction. The walls separating the jet from the coflows have a thickness of $L/28$, with $L=\SI{5}{mm}$ being the width of the central jet. The mesh is uniform in streamwise and spanwise directions. In the crosswise direction, the mesh is uniform in the center between $-L \leq y \leq L$ and coarsened with a linear profile in the coflow region.
The mesh resolution $\Delta = \SI{30}{\mu m}$ is chosen to adequately resolve the Kolmogorov length scale as well as the flame front. The flame is resolved with 15 cells within the thermal flame thickness. The adopted resolution leads to a fully resolved simulation (FRS) where no compromises are present in the description of the flame structure~\cite{domingo_2023}.
A comprehensive listing of all simulation parameters and characteristic numbers is given in \tablename\ref{tab:characteristic_scales}.

The unburnt hydrogen-air mixture is chosen according to the observation of strong thermo-diffusive instabilities in laminar conditions. The equivalence ratio is set to $\phi=0.5$, the pressure is $p=\SI{1}{atm}$ and the temperature of the central jet is specified as $T=\SI{300}{K}$.
At these conditions, the reference laminar flame speed and the laminar thermal flame thickness of a planar unstretched flame is $s_\mathrm{L} = \SI{0.517}{m/s}$ and $\delta_\mathrm{L} = \SI{0.445}{mm}$, respectively.
The thermal flame thickness is defined by $\delta_\mathrm{L} = (T_\mathrm{b} - T_\mathrm{u})/\mathrm{max}(\nabla T)$, with $T_\mathrm{b}$ and $T_\mathrm{u}$ being the burnt and unburnt temperature, respectively, and $\nabla T$ the temperature gradient of a one-dimensional unstretched flame.
The coflow composition corresponds to the burnt gas state of the central jet's mixture and the temperature of the burner walls is fixed at $T_\mathrm{wall} = \SI{300}{K}$.
The Reynolds number of the central jet is $\mathrm{Re} = U_\mathrm{jet}L / \nu = 10,000$, with $U_\mathrm{jet}$ being the bulk velocity, and the kinematic viscosity of the unburnt mixture $\nu$.
The coflows are laminar with an inlet velocity $U_\mathrm{coflow} = \SI{9.5}{m/s}$.
The Kolmogorov length scale $\eta$ is determined at the position of the mean flame sheet and at half the height of the mean flame length. The location inside the mean flame sheet is selected at a temperature $T = \SI{970}{K}$ corresponding to the mean between the fresh and burnt gas temperature in a one-dimensional unstretched flame of identical composition.
For the investigated operating condition, a moderate Karlovitz number $\mathrm{Ka} = (\delta_\mathrm{L}/\eta)^2 \approx 30$ is obtained.

The velocity data for the central jet is obtained from a precursor simulation of an inert, fully developed turbulent channel flow. 
In spanwise direction, periodic boundary conditions are employed.
All outlets are modeled with non-reflective boundary conditions.
    
\begin{table}[ht]
    \centering
    \footnotesize
    \caption{Parameters and characteristic numbers of the simulation.} 
    \begin{tabular}{cc}
        \toprule
        $L$ [mm] & 5  \\
        $U_\text{jet}$ [m/s] & 38 \\
        $U_\text{coflow}$ [m/s] & 9.5 \\
        $\eta$ [$\mu$m] & 80 \\
        $\Delta$ [$\mu$m] & 30 \\
        $s_\mathrm{L}$ [m/s] & 0.517 \\
        $\delta_\mathrm{L}$ [mm] & 0.445 \\
        $\mathrm{Re}$ & 10,000 \\
        $\mathrm{Ka}$ & 30 \\
        ($L_{x}$, $L_{y}$, $L_{z}$)$/L$ & 12, 12, 2 \\
        $N_{x}$, $N_{y}$, $N_{z}$ & 2000, 510, 333 \\
        \bottomrule
    \end{tabular}
    \label{tab:characteristic_scales}
\end{table}

\subsection{Governing Equations and Numerical Methods}
The simulation is performed using the finite volume method framework OpenFOAM~\cite{weller_1998, morev_2022} solving the compressible reactive Navier-Stokes equations.
Second-order discretization schemes are employed in time and space. The methods have been validated for performing accurate FRS of reactive flows~\cite{zirwes_2023b}.
Chemical reactions are modeled by directly solving the chemical source terms employing the detailed reaction mechanism by Li et al.~\cite{li_2004} containing 9 species and 21 reactions. The species diffusivities are modeled according to the mixture-averaged transport approach by Curtiss and Hirschfelder~\cite{curtiss_1949}.

\subsection{Composition space model (CSM)} \label{sec:CSM}
In the remainder of the manuscript, a composition space model~(CSM)~\cite{scholtissek_2019a,scholtissek_2019b,boettler_2021,xychen_2021} is used to model differential diffusion and stretch effects in a flamelet-based manner.
In the CSM, conservation equations for species mass fractions $Y_k$, temperature $T$ and progress variable gradient $g_c=|\nabla Y_c|$ (required for closure) are solved along progress variable $Y_c$ which spans the composition space. 
The system of equations reads:

\begin{align}
   \rho\dd{Y_k}{\tau}
   &= - \underbrace{g_c\dd{}{Y_c}\left(g_c\,\rho Y_k\tV_k\right)
   + g_c\dd{}{Y_c}\left(g_c\,\rho Y_c\tV_c\right)\dd{Y_k}{Y_c}}_{\text{Diffusion}} \notag\\
   &+\underbrace{\rho g_c\,\kappa_c\left(Y_k \tV_k - Y_c\tV_c\dd{Y_k}{Y_c}\right)}_{\text{Diffusion}}
   -\underbrace{\dot{\omega}_c\dd{Y_k}{Y_c}}_{\text{Drift}} + \underbrace{\rho \dot{\omega}_k}_{\text{Source}} \,,
  \label{eq:Y_pflamelet} \\
  \rho\dd{T}{\tau}
  &= \underbrace{\frac{g_c}{c_p}\dd{}{Y_c}\left(g_c\lambda\dd{T}{Y_c}\right)
  +g_c\dd{}{Y_c}\left(g_c\,\rho Y_c\tV_c\right)\dd{T}{Y_c}}_{\text{Diffusion}}\notag\\
   &- \underbrace{\rho g_c^2  \sum_k^{n_s}\frac{c_{p,k}}{c_p} Y_k \tV_k\dd{T}{Y_c}
   -\rho g_c\,\kappa_c\left(\frac{\lambda}{\rho c_p} + Y_c\tV_c\right)\dd{T}{Y_c}}_{\text{Diffusion}}\notag\\
    &- \underbrace{\dot{\omega}_c\dd{T}{Y_c}}_{\text{Drift}} + \underbrace{\frac{\dot{\omega}_T}{c_p}}_{\text{Source}} \,,
  \label{eq:T_pflamelet} \\
  0 &= - \underbrace{g_c^2 \ddtwo{}{Y_c}\left(g_c\,\rho Y_c\tV_c\right)
  +g_c^2 \dd{}{Y_c}\left(\kappa_c\,\rho Y_c\tV_c\right)}_{\text{Diffusion}}\notag\\
  &- \underbrace{\dot{\omega}_c\dd{g_c}{Y_c}}_{\text{Drift}} + \underbrace{g_c \dd{\dot{\omega}_c}{Y_c}+ \rho K_s g_c}_{\text{Source}} \,,
  \label{eq:gc_pflamelet}
\end{align}

with the diffusion velocity of species $k$ with respect to the composition space $\tV_k$, the heat conductivity $\lambda$, the heat capacity $c_p$ and the heat release rate $\dot{\omega}_T$.
The different terms in this set of equations can be classified as diffusion, drift, and source terms. 
As diffusion and source terms are well-established and known from conventional transport equations, we will not elaborate on them here for brevity.
However, the drift terms are less common. These terms scale the composition space and are crucial for the CSM framework as the burned side boundary condition is coupled to the progress variable gradient which subsequently scales the computational domain.
This is an important feature of these equations as stretched flame structures can be studied by supplying strain $K_s=\nabla_t\cdot\mbf{u}_t - \left(\mbf{u}\cdot\mbf{n_c}\right)\kappa_c$ and curvature $\kappa_c = -\nabla\cdot\mbf{n}_c$ as external parameters.
It is well known that depending on the overall flame stretch (negative or positive) sub- or super-equilibrium states can be reached which coincide with different burned side values for the progress variable. The drift terms ensure that this behaviour is accurately described and the appropriate burned side value of the progress variable is obtained by the CSM.
By that, the characteristics of various canonical flame configurations are reproduced by using only one set of equations. Further, the CSM has a larger attainable strain-curvature parameter space compared to canonical configurations and flame structures for arbitrary combinations of strain and curvatures can be studied~\cite{boettler_2021}.
In particular, negative curvatures and strain rates can be considered for further analysis. 
This is important for hydrogen-air flames due to their strong sensitivity to stretch effects. It was found in our previous work that different contributions of strain and curvature to the overall flame stretch alters the flame characteristics significantly.
The interested reader is referred to~\cite{boettler_2021,xychen_2021}, where a detailed validation against several canonical flame configurations and a systematic analysis of arbitrary combinations of strain and curvature are provided. 
Further, the usage of the CSM to study stretch effects is also expected to be beneficial for turbulent premixed hydrogen-air flames as it was reported by Amato et al.~\cite{amato_2015} that the correlation between strain and curvature is reverted between laminar canonical flames and turbulent flames which raises the need of developing extended laminar model configurations.
Note that the progress variable definition in all calculations is $Y_c = Y_{\text{H2O}} - Y_{\text{H2}} - Y_{\text{O2}}$.

\section{Manifold generation}
Next, stretch effects and their corresponding thermochemical states are analyzed in the FRS of the turbulent slot flame. Thereby, the parameter ranges for strain and curvature are estimated, which need to be taken into account for the subsequent flamelet modeling based on the CSM.
Further, an optimal estimator assessment is performed to confine potential sets of parameters that can be used to parameterize the manifold and to investigate the modeling error of certain parameter reductions.
Finally, various flamelet manifolds are introduced which account for different physical effects and are generated based on different parameter variations using the CSM.

\subsection{Analysis of stretch effects for the turbulent flame} \addvspace{5pt}

While overall flame stretch effects are widely discussed in the literature, this study focuses on contributions of strain $K_s$ and curvature $\kappa_c$ effects to the flame stretch using following stretch decomposition~\cite{groot_2002a}:
\begin{equation}
      K = \underbrace{\nabla_\mathrm{t}\cdot\mbf{u}_\mathrm{t} - \left(\mbf{u}\cdot\mbf{n}_c\right)\kappa_c}_{K_s}
  + s_\mathrm{d}\,\kappa_c \,,
\end{equation}
where $s_\mathrm{d}$ represents the flame displacement speed.

\begin{figure}[ht]
\centering
\includegraphics[scale=1.1]{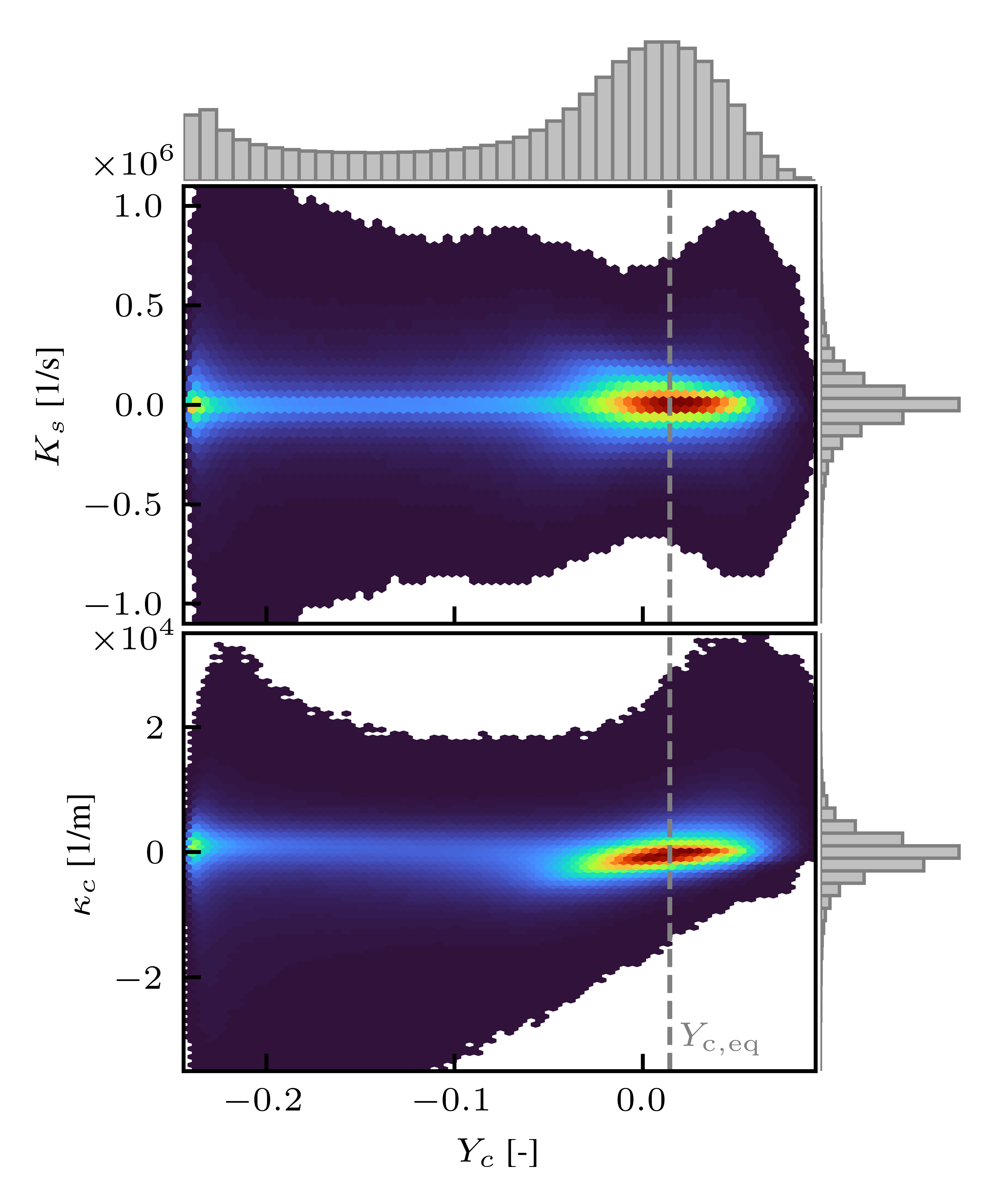}
\caption{Joint probability density functions (PDFs) of strain $K_s$ and progress variable $Y_c$ (top) and the one of curvature $\kappa_c$ and progress variable (bottom) with marginal PDFs shown as bars at the respective edges. The color coding of the scatter indicates the likelihood of certain states starting from unlikely (blue) to most probable states (red). The dashed grey line corresponds to the equilibrium progress variable value $Y_\mathrm{c,eq}$ of an unstretched planar flame with an equivalence ratio of $\phi=0.5$.}
\vspace{-4pt}
\label{fig:jpdf_strain_curvature}
\end{figure}

In \figurename~\ref{fig:jpdf_strain_curvature}, the joint probability density functions~(PDFs) of strain and progress variable (top) and the one of curvature and progress variable (bottom) of the turbulent flame are depicted. 
For visual inspection, marginal PDFs are depicted at the respective axes and the equilibrium progress variable value, $Y_\mathrm{c,eq}$, for an unstretched planar hydrogen-air flame (equivalence ratio of $\phi=0.5$) is shown as a grey dashed line. Both quantities ($K_s$, $\kappa_c$) show a broad variation in the turbulent flame configuration. The marginal PDF for strain is symmetrical, while the one for curvature is slightly skewed toward negative values. 
This behavior is expected in turbulent hydrogen-air flames and in agreement with the literature as it was reported that hydrogen addition shifts the curvature distribution to negative values~\cite{renou_2000, fru_2012, cecere_2016}.
The qualitative PDF of the progress variable has a bimodal shape where a first peak occurs close to the unburned side and a second one at the burned side where the progress variable is close to the equilibrium value of the corresponding unstretched planar flame. Further, a broad scatter exists around the burned side peak. 
The joint PDF of strain and progress variable shows an overall symmetrical distribution between positive and negative strain rates where the strain is in the order of magnitude of $10^6\, \text{m}^{-1}$ while approximately 50\% of the values are between $\approx \pm$\SI{70000}{\per \second}. 
However, the joint PDF of curvature and progress variable exhibits a skewed distribution. The curvature reaches values up to $\approx \pm$\SI{30000}{\per\meter} but 90 \% of the data is located in the range of $\SI{-7000}{\per\meter} \leq \kappa_c \leq \SI{5500}{\per\meter}$. Positive curvatures are found for both low and high $Y_c$ values, while negative curvatures are predominantly found close to the unburned side and only rarely result in super equilibrium states ($Y_c > Y_\mathrm{c,eq}$). 

These observations indicate a link between turbulent fluctuations and the flame response. The strain distribution is symmetric since the imposed flow field does not alter the flame front in any predominant direction. However, the curvature distribution is skewed due to the flame's response. Positive curvature promotes the flame propagation and causes super equilibrium states while negative curvature weakens the flame and subsequently, sub-equilibrium states ($Y_c < Y_\mathrm{c,eq}$) are found. 
This correlation of positive curvature with super-equilibrium states coincides with variations in the local flame structure and subsequently with altered burning rates due to the Lewis number effect. This effect was already discussed in the context of laminar thermo-diffusively unstable flames and needs to be further elaborated in turbulent flames~\cite{boettler_2022b, berger_2019,  howarth_2022}.

\begin{figure}[ht]
\centering
\includegraphics[width=70mm]{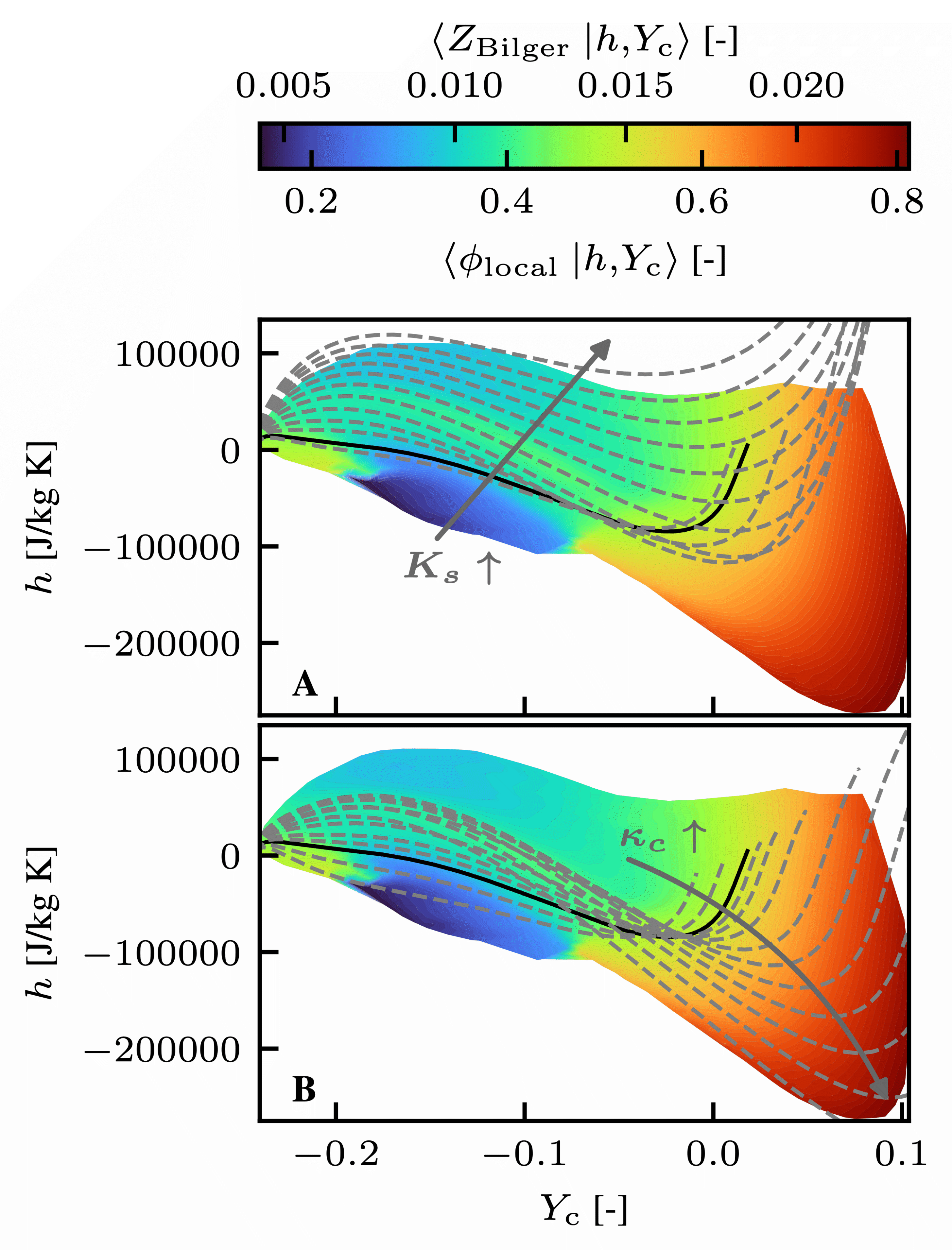}
\caption{Conditional mean of Bilger mixture fraction $Z_\mathrm{Bilger}$ and local equivalence ratio $\phi_\mathrm{local}$ conditioned on enthalpy $h$ and progress variable $Y_c$. Additionally, CSM solutions with varying strain ($\SI{-100}{\per\second} \leq K_s \leq \SI{28000}{\per\second}$;  $\kappa_c=\SI{0}{\per\meter}$) are depicted as dashed grey lines in the upper figure (A). In the bottom figure (B), the dashed grey lines correspond to CSM simulations with varying curvature ($\SI{-500}{\per\meter} \leq \kappa_c \leq \SI{8000}{\per\meter}$;  $K_s=\SI{0}{\per\second}$). The solid black line corresponds to the unstretched planar flame solution. All CSM results follow the global conditions of the turbulent flame~($T=\SI{300}{K}, p=\SI{1}{atm}, \phi=0.5$).}
\label{fig:conditional_statistics}
\end{figure}

Since the overall distributions of strain and curvature can already be related to the flame response, their impact on the thermochemical states in the turbulent flame is of interest. Therefore, the thermochemical states are depicted in composition space in \figurename\ref{fig:conditional_statistics}. 
This figure shows the scatter of thermochemical states spanned by the enthalpy $h$ and progress variable $Y_c$ and is colored by the conditional mean of the Bilger mixture fraction~\cite{bilger_1990}, which can also be interpreted as a local equivalence ratio $\phi_\mathrm{local}$. 
In addition, CSM simulations with varying strain ($\SI{-100}{\per\second} \leq K_s \leq \SI{28000}{\per\second}$;  $\kappa_c=\SI{0}{\per\meter}$) are depicted as grey dashed lines in \figurename\ref{fig:conditional_statistics}~A and CSM results with varying curvature ($\SI{-500}{\per\meter} \leq \kappa_c \leq \SI{8000}{\per\meter}$; $K_s=\SI{0}{\per\second}$) are shown similarly in \figurename\ref{fig:conditional_statistics}~B. The solid black line corresponds to the unstretched planar flame solution, respectively.

The thermochemical states in the turbulent flame scatter broadly concerning the enthalpy and the Bilger mixture fraction along the progress variable. 
It is noted that the enthalpy is a suitable parameter to judge the composition space since it includes the influence of all species.
For low progress variable values, the enthalpy varies significantly with higher mixture fraction values located in the center of the scatter and leaner mixtures (low values of the Bilger mixture fraction) are found towards the respective edges of the scatter (lower and higher enthalpy levels). However, for high values of the progress variable, the enthalpy scatter broadens even further and significantly lower enthalpy levels are found, depicting also substantially higher mixture fraction values.
When comparing the scatter of the turbulent flame against the CSM calculations with varying strain or curvature, the predominant influences of strain and curvature become visible. For higher enthalpy levels, the scatter of the turbulent flame is well captured by CSM simulations with increasing strain. It is highlighted that not only a comparable parameter space is spanned, but also the profiles of the CSM resemble the overall shape of the turbulent scatter~(\figurename\ref{fig:conditional_statistics}~A). Similarly, for high values of progress variable and reduced levels of enthalpy, the range and shape of the scatter of the turbulent flame simulation are captured by CSM simulations with varying curvature~(\figurename\ref{fig:conditional_statistics}~B). Only a small part of the scatter where the lowest values of Bilger mixture fraction occur are not directly captured by the CSM results. This can be either attributed to a joint interplay of strain and curvature, or multidimensional effects. It is noted that multidimensional effects cannot be captured by the CSM since these states correspond to highly negative curvatures where flame-tangential diffusion tends to become increasingly important~\cite{boettler_2022b}. 
Remarkably, the CSM calculations capture major parts of the scatter of the turbulent flame even though nominal values for strain and curvature in the FRS significantly exceed the strain-curvature parameter space attainable with the CSM. Additionally, these findings indicate that both strain and curvature effects are important when modeling these types of flames. In the following, the relevance of these effects is further elaborated. 

\subsection{Optimal estimator assessment}\label{sec:opt_est}

Next, a suitable parameterization must be found to enable the construction of a suitable manifold which accounts for all relevant physical phenomena. Therefore, an optimal estimator assessment is carried out for different parameterizations~\cite{moreau_2006,berger_2022}. In an optimal estimator assessment, a set of parameters~$\psi$ is used to parameterize a quantity (e.g. the progress variable source term~$\dot{\omega}_c$) and is determined by an error norm known as irreducible error.
Note that the progress variable source term $\dot{\omega}_c$ is used in this study since it represents an important characteristic quantity that needs to be predicted by manifold-based models and properly capturing the source term in the manifold poses usually one of the biggest challenges in flamelet-based models. 
Further, this error is quantified by the quadratic error norm of the scatter of $\dot{\omega}_c$ relative to the conditional mean $\langle\dot{\omega}_c|\psi\rangle$. Subsequently, this error is normalized by the maximum obtained from a respective unstretched flamelet~\cite{berger_2022}:
\begin{equation}
\epsilon^\mathrm{norm}_{\mathrm{irr},\dot{\omega}_c} = \frac{\langle (\dot{\omega}_c - \langle \dot{\omega}_c | \psi \rangle)^2 | Y_c \rangle}{\mathrm{max}(\dot{\omega}_c^\mathrm{flamelet})^2}    
\end{equation}
Thereby, low values for the normalized irreducible error $\epsilon^\mathrm{norm}_{\mathrm{irr},\dot{\omega}_c}$ indicate a suitable parameterization of the thermochemical states by the respective set of parameters~$\psi$. 

\begin{figure}[ht]
\centering
\includegraphics[scale=1.0]{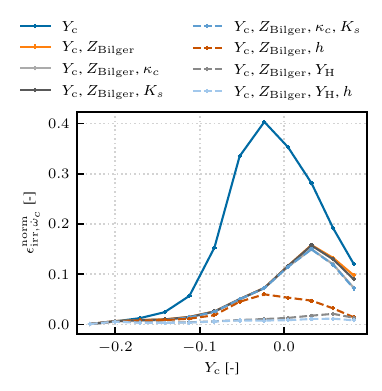}
\caption{Irreducible errors of $\dot{\omega}_c$ for different sets of parameters.}
\label{fig:optimal_estimator}
\end{figure}

In \figurename\ref{fig:optimal_estimator}, the irreducible error is shown for different combinations of parameters along the progress variable.
High errors are obtained when parameterizing only by the progress variable $Y_c$. The irreducible error is reduced when adding the Bilger mixture fraction to the set of parameters. Including strain $K_s$, curvature $\kappa_c$, or both in the parameterization does not result in a significant error reduction. This is attributed to the fact that strain and curvature are highly instantaneous quantities that are closely linked to the flame's topology. However, they do not account for the temporal evolution of the thermochemical state~\cite{klimenko_2002}. Zirwes et al.~\cite{zirwes_2021,zirwes_2023} found a phase shift between stretch effects and the subsequent flame response, which depends on the local flow time scale and discussed memory effects in the local flame dynamics.
Such memory effects could also play a role here. Thus, even though it was shown in \figurename\ref{fig:conditional_statistics} that the flame parameters strain and curvature induce the desired changes in the flame structure, they are not necessarily suitable control variables for manifold parameterization.
However, strain and curvature effects can be represented by respective characteristic quantities to avoid any phase shift in the thermochemical states. The H-radical mass fraction~$Y_\text{\ce{H}}$ was already used to account for curvature effects in our previous works~\cite{boettler_2022b,wen_2021a}.

Further, when revisiting the discussion concerning \figurename\ref{fig:conditional_statistics}, the enthalpy was found to be suitable to distinguish between strain and curvature effects, since CSM calculations with varying strain or curvature cover different parts of the state space of the turbulent slot flame.
Utilizing $Y_\text{H}$ or $h$ as markers for strain and curvature results in an additional error reduction. The lowest irreducible error and, hence, the most suitable parameterization is obtained when using $\psi(Y_c, Z_\mathrm{Bilger}, Y_\text{\ce{H}}, h)$. 
This combination of the number of scalars and obtained error is considered a good tradeoff between numerical cost (manifold size, scalars to be solved) and accuracy.

\subsection{Generation of flamelet manifolds}

Based on the findings of the optimal estimator assessment various flamelet-based manifolds are generated. Thereby, not only their parameterization but also the characteristics of the one-dimensional flamelets used for the manifold generation are varied.
In this work, four different manifolds, which are generated from different CSM data sets and parameterized by different control variables, are used including a novel manifold based on machine learning. Thereby, the aim is to subsequently address modeling errors resulting from unstretched and stretched flamelets.
The manifold parameterization follows the conceptual approach of our previous work where the manifold is coupled to the CFD by solving transport equations for the major species (\ce{H2}, \ce{O2}, \ce{H2O}) instead of $Z_\mathrm{Bilger}$ and $Y_c$~\cite{boettler_2022a,boettler_2022b}.

The first manifold (M-1) is based on unstretched planar CSM calculations with varying equivalence ratio~($0.275 \leq \phi \leq 0.85$) and is parameterized by the Bilger mixture fraction $Z_\mathrm{Bilger}$~\cite{bilger_1990} and progress variable $Y_c$. 
This manifold parameterization is well established and frequently used~\cite{regele_2013, schlup_2019} and captures mixture stratification.

The second manifold (M-2) is a subsequent extension of the first manifold, including also CSM calculations at different enthalpy $h$ levels (preheating and heat loss effects) and therefore parameterized by $Z_\mathrm{Bilger}$, $h$ and $Y_c$. This parameterization is more complex than the one of M-1 but was already used in several studies~\cite{mukundakumar_2021,boettler_2022a} and captures both mixture stratification and non-adiabatic effects. It is noted that differential diffusion effects are captured by the Bilger mixture fraction $Z_\mathrm{Bilger}$, but they are also partly manifested in the enthalpy. Therefore, considering enthalpy as control variable can be beneficial for the manifold parameterization which is already indicated in the previous section.

The third manifold (M-3) is based on CSM calculations with different equivalence ratios~(c.f. M-1) and varying curvature~($\SI{-2500}{\per\meter}\leq \kappa_c \leq \SI{8000}{\per\meter}$). It is parameterized by $Z_\mathrm{Bilger}$, $Y_c$ and the H radical mass fraction $Y_\mathrm{H}$. The latter one correlates with the curvature variation and was already utilized to parameterize a manifold addressing laminar thermo-diffusively unstable hydrogen-air flames~\cite{boettler_2022b}. This manifold parameterization is even more challenging concerning its generation and was only used in our previous work since curvature can be supplied to the CSM as an external parameter~\cite{boettler_2022b}. It should be noted that one-dimensional flames with negative curvature cannot be easily addressed by flamelet solvers operating in physical space.

An additional novel manifold (M-4) is developed in this study to assess the joint influence of strain and curvature.
The data used for M-4 is an extension of the CSM solutions used for M-3 and further includes a strain variation in the range of $\SI{-400}{1/s} \leq K_s \leq \SI{30000}{1/s}$ as an additional dimension. 
It is noted that the thermochemical states are retrieved by linear interpolation in the case of M-1, M-2 and M-3, while M-4 is generated as an artificial neural network, to reduce the overall memory footprint of such high dimensional manifolds including also differential diffusion effects. A review of data-driven models in combustion research can be found in~\cite{ihme_2022}.
M-4 is parameterized analogously to the other manifolds using the same approach but utilizing the major species (\ce{H2}, \ce{O2}, \ce{H2O}) directly as model inputs to approximate $Z_\mathrm{Bilger}$ and $Y_c$.
Thereby, the model inputs are the species mass fractions of \ce{H2}, \ce{O2}, \ce{H2O}, \ce{H}, and enthalpy~$h$. These control variables are found to be the most suitable to capture the effects in the turbulent flame configuration, as already shown by the optimal estimator assessment provided in \figurename\ref{fig:optimal_estimator}. 
It is noted that the accuracy of both manifold approaches where verified for manifolds M-1, M-2 and M-3 and a comparable accuracy of both approaches is found. 

\begin{table}
    \centering
    \caption{Overview of utilized manifolds classified by manifold name, parameterization of the manifold data and the set of control variables to access the respective manifold.} 
    \footnotesize
    \begin{tabular}{cccc}

        \toprule
        Manifold name & Manifold data & Number of CSM calculations & Control variables \\\toprule
        M-1 & $\psi(\phi,Y_c)$ & 30 & $Z_\mathrm{Bilger}, Y_c$ \\
        M-2 & $\psi(\phi,h,Y_c)$ & 750 & $Z_\mathrm{Bilger},h, Y_c$ \\
        M-3 & $\psi(\phi,Y_c,\kappa_c)$& 1140 & $Z_\mathrm{Bilger}, Y_c, Y_\mathrm{H}$ \\
        M-4 & $\psi(\phi,Y_c,\kappa_c, K_s)$& 26350 & $Y_\mathrm{\ce{H2}}, Y_\mathrm{\ce{O2}}, Y_\mathrm{\ce{H2O}},$ $Y_\mathrm{H}, h$ \\\bottomrule
    \end{tabular}
    \label{tab:models}
\end{table}

An overview of the manifolds used in this study is given in \tablename~\ref{tab:models}. The manifold name is shown together with the parameterization of the manifold data, which represents the thermochemical states $\psi$ spanned by various CSM simulations, the number of CSM calculations included in the parameter variation and the set of control variables that are used to access the manifold. Note that the progress variable definition in all manifolds is $Y_c = Y_{\text{H2O}} - Y_{\text{H2}} - Y_{\text{O2}}$ and the overall computational cost to generate the manifolds increases with the number of CSM calculations.

\section{\textit{A-priori} analysis}
\begin{figure*}[hbt]
\centering
\includegraphics[scale=1]{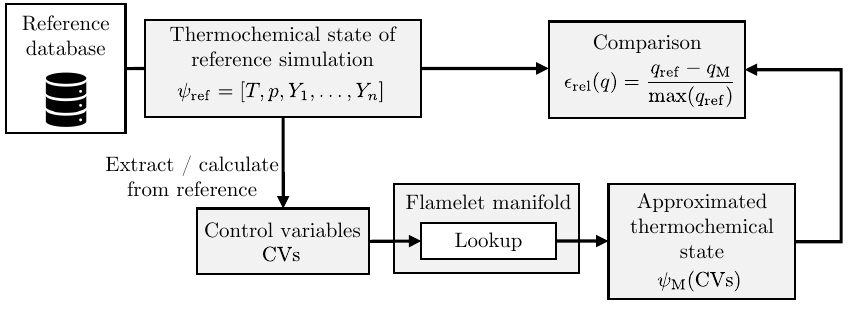}
\caption{Schematic overview of an \textit{a-priori} analysis. Adapted from~\cite{steinhausen_2023}.}
\label{fig:apriori_schematic}
\end{figure*}

An \textit{a-priori} analysis is performed to assess the general capabilities of the four manifolds to capture the microstructure of the lean turbulent hydrogen-air flame.
In an \textit{a-priori} analysis a reference database is utilized to quantify the performance of flamelet manifolds. The overall procedure is depicted schematically in \figurename\ref{fig:apriori_schematic}. In this study the full resolved simulation of the turbulent slot flame acts as reference data based on which the reduced set of control variables of the flamelet manifolds is extracted or calculated. Subsequently, these control variables are used to perform a lookup of the flamelet manifold to retrieve an approximated thermochemical state from the manifold. This approximated thermochemical state is compared to the reference data by computing a relative error $\epsilon_\mathrm{rel}$ between the reference data and the manifold prediction of a quantity $q$. 

\begin{figure*}[p]
\centering
\includegraphics[scale=0.98, angle=90]{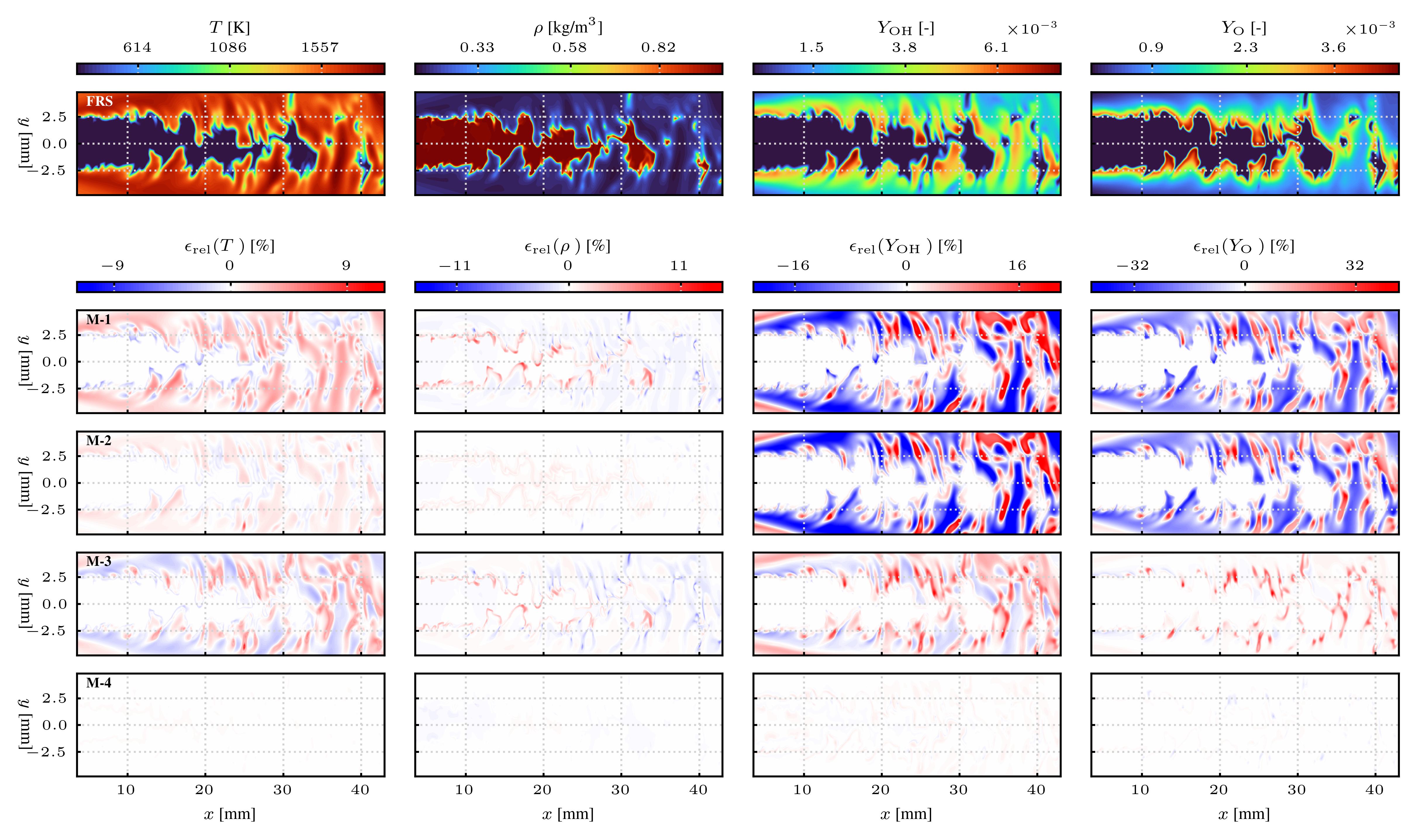}
\caption{\textit{A-priori} comparison of the manifolds against the fully resolved reference solution for an instantaneous snapshot. The temperature $T$, density $\rho$ and the species mass fractions of \ce{OH} and \ce{O} fields of the reference data are shown along the $x,y$-coordinates and $z=\SI{0}{mm}$ (top row). The manifold predictions are depicted as relative deviations 
for the respective fields.}
\label{fig:apriori}
\end{figure*}

\begin{figure*}[p]
\centering
\includegraphics[scale=0.99, angle=90]{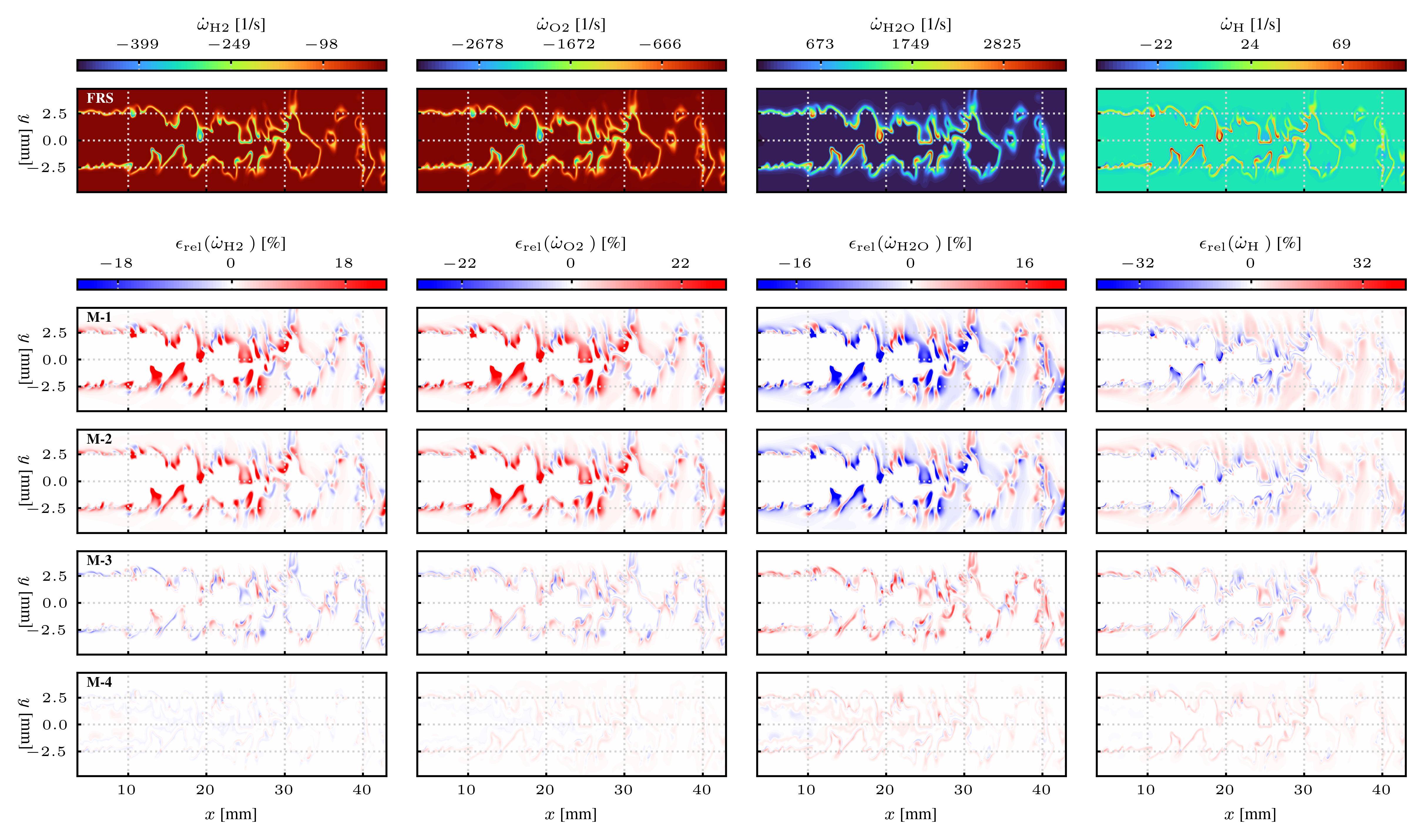}
\caption{\textit{A-priori} comparison of the manifolds against the fully resolved reference solution for an instantaneous snapshot. The species source terms of \ce{H2}, \ce{O2}, \ce{H2O} and \ce{H} of the reference data are shown along the $x,y$-coordinates and $z=\SI{0}{mm}$ (top row). The manifold predictions are depicted as relative deviations for the respective fields.}
\label{fig:apriori_reaction_rates}
\end{figure*}

In \figurename\ref{fig:apriori}, a snapshot of the temperature $T$, density $\rho$ and the species mass fractions of \ce{OH} and \ce{O} fields of the turbulent flame are shown along the $x,y$-coordinates and $z=\SI{0}{mm}$ (top row). Further, the relative deviations of all manifolds are depicted for the respective fields.
The turbulent flame front is highly curved and various pockets containing unburned mixture are observed, which are detached from the main flame front and can be found for $x>\SI{30}{mm}$.
Maximum temperatures higher than the equilibrium temperature of an unstretched flame are found near the flame front, indicating super equilibrium states caused by the interplay of differential diffusion and flame stretch. 
Further, this region is intersected by areas of lower temperatures as a consequence of turbulent fluctuations. The density and radical mass fraction fields show similar trends as the temperature field. Besides the impact of local temperature fluctuations, the amount of \ce{OH} and \ce{O} mass fractions also depend significantly on curvature. In areas convexly shaped towards the burned side (positive curvature), increased amounts of radicals are found, highlighting the increased burning rates due to the Lewis number effect.
These characteristics highlight the relevance of differential diffusion effects in this flame configuration and indicate the interplay of turbulent wrinkling and thermo-diffusive instabilities. 
The prediction of M-1 and M-3 show deviations in the temperature field up to \SI{10}{\percent}. In particular, the temperatures close to positively curved flame segments are significantly underpredicted. The same observation applies to the density field. M-2 exhibits better agreement with the reference data for $T$ and $\rho$, showing the highest deviations around the detached pockets of the unburned mixture. This improvement is expected since M-2 uses enthalpy as a model input and thermochemical states corresponding to preheated unstretched flames are accessed. However, the tabulation approaches based on unstretched flames cannot predict the radical species in this flame configuration. M-1 and M-2 depict high deviations for the \ce{OH} and \ce{O} radical mass fractions. M-3, which accounts for curvature effects, gives better predictions for the radical mass fractions. The highest deviations (up to $20-\SI{25}{\percent}$) are located close to negatively curved flame segments. Only the prediction of M-4, taking both strain and curvature effects into account, results in negligible errors for all quantities. 

Besides an adequate description of the flame structure, predicting the source terms of the control variables is of utmost importance since they determine the overall transport and flame dynamics in coupled simulations. Following the coupling approach of transporting major species (as in our previous work~\cite{boettler_2022a,boettler_2022b,lou_2022}) the source terms of $Y_{\text{H2}}, Y_{\text{O2}}, Y_{\text{H2O}}, Y_{\text{H}}$ are required. These source terms are shown in \figurename\ref{fig:apriori_reaction_rates} together with the respective relative error of the manifold predictions.
The source terms of the major species follow the previously discussed Lewis number effect and their variation also indicate the locally altered burning rate of the flame.
The highest source terms are found in flame segments with positive curvature. Note that this corresponds to negative source terms for \ce{H2} and \ce{O2} since these species get consumed. Negatively curved flame segments depict lower source terms. The source term of the H radical shows the highest production in positively curved flame segments. However, also strong consumption takes place in close vicinity of these areas. In contrast, only small source terms are found in negatively curved reaction zones.
The manifold predictions of M-1 and M-2 depict errors beyond \SI{50}{\percent} for all source terms. In particular, the source terms are significantly underpredicted in positively curved flame segments. M-3 shows reduced but significant errors in the range of \SI{20}{\percent}. 
However, the M-4 manifold recovers the source terms of the reference data with almost negligible errors for the major species. Only small deviations are observed for the H radical source term.

\begin{figure*}[hbt]
\centering
\includegraphics[scale=1.0]{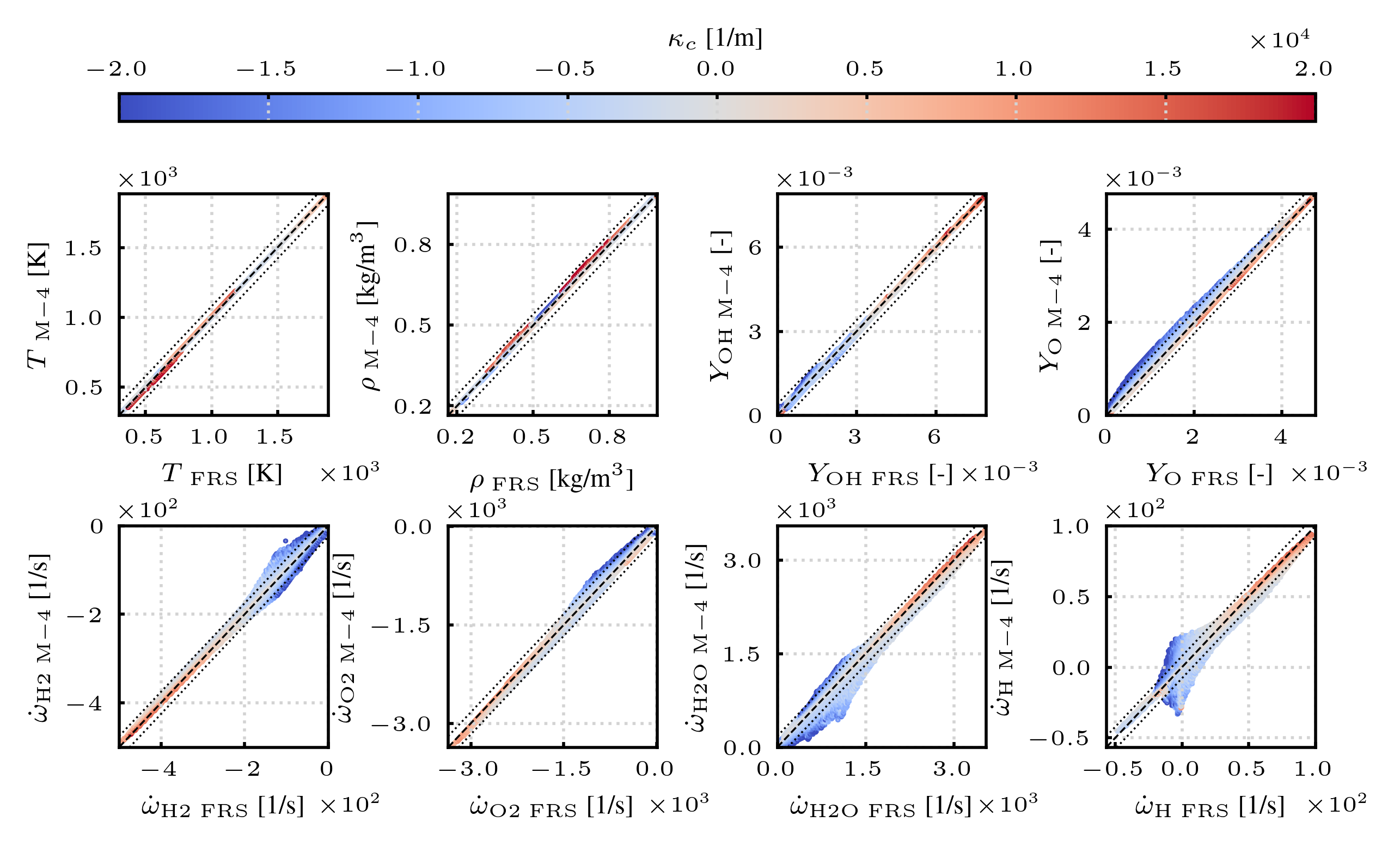}
\caption{Correlation of reference data and the M-4 manifold prediction for various characteristic quantities utilizing the three-dimensional data set. A perfect correlation is highlighted by the black dashed line and the dotted lines represent a 5\% error margin based on the absolute variation found in the fields of the reference data, respectively.}
\label{fig:apriori_correlation}
\end{figure*}

Finally, the predictions of the M-4 manifold are investigated in more detail since they show the best agreement with the reference data. In \figurename\ref{fig:apriori_correlation}, the previously discussed fields of the turbulent flame reference data are correlated with the respective manifold prediction, which allows for a more detailed local error assessment on the three-dimensional data set.
Further, the scatter is color-coded by curvature. A full correlation of the manifold prediction and the reference data is indicated by the black dashed line. Further, the dotted lines correspond to an error margin of $\pm 5$~\%. 
Very good agreement is found for the temperature $T$ and density $\rho$ prediction. The correlation is slightly weaker for the radical mass fractions of \ce{OH} and \ce{O}, where the scatter broadens slightly for points corresponding to negative curvature.
Nevertheless, the deviations for the \ce{OH} radical mass fraction do not exceed the 5 \% error margin. Only the prediction of the \ce{O} mass fraction slightly exceeds the error margin for points corresponding to negatively curved flame segments. A similar behavior is found for the source terms of the major species. The weakest correlation is obtained for predicting the H radical source term~$\dot{\omega}_\text{H}$. Overall, the manifold prediction stays mostly within the 5 \% error margin. However, points with negative curvature show higher deviations. Under these conditions, the H radical source term scatters around zero. Therefore, these errors are considered insignificant since they are topologically confined and occur in areas of weak reactions.
Based on this detailed error estimation of different manifolds, it is concluded that manifold-based models based on unstretched planar flames cannot predict the thermochemical states of the lean turbulent hydrogen-air flame while including only curvature effects in the manifold leads to mediocre results. Only if both positive and negative values for strain and curvature are considered in the manifold generation good agreement with the reference data is achieved.

\section{Summary and conclusions} \addvspace{10pt}
In this work, a fully resolved simulation of a turbulent lean premixed hydrogen-air slot flame is performed using detailed chemistry. This flame configuration is analyzed concerning stretch effects and serves as a reference database for developing flamelet-based models. 
The performance of various tabulated manifolds in predicting the thermochemical states in the turbulent flame is assessed in an \textit{a-priori} manner. 

The analysis of strain and curvature in the turbulent flame reveals that these effects not only alter the mixture composition but also notably influence the enthalpy.
Additionally, it is shown that major parts of the scatter of thermochemical states found in the turbulent reference simulation can be captured by one-dimensional calculations with a composition space model (CSM) separately varying strain and curvature. 
CSM calculations with increasing strain result in higher enthalpy levels with moderate shifts in local mixture composition, while CSM calculations with increasing curvature lead to overall lower enthalpy levels and significantly richer mixtures.

Based on these findings, four different flamelet manifolds with increasing complexity are generated, including (1)~a manifold generated from adiabatic unstretched flames, (2)~a manifold based on non-adiabatic unstretched flames, (3)~a manifold accounting for curvature variations, and (4)~a novel manifold considering strained and curved flames. The manifold predictions are discussed concerning their capability to describe the structure of the turbulent flame. It is shown that significant modeling errors are obtained when using unstretched planar flames to generate the manifold.
Although including enthalpy variations in these flamelets leads to an improved temperature field prediction, significant deviations persist for species fields and source terms, indicating that these modeling approaches cannot capture the synergistic effects of turbulence and thermo-diffusive instabilities.
The deviations are reduced when considering CSM calculations with varying curvature. The prediction of the manifold, taking positive and negative strain and curvature into account, shows the best agreement with the reference data. Hence, the altered reaction intensity caused by combined stretch and Lewis number effects is well recovered by the novel manifold.
These results highlight the importance of stretch effects in tabulated manifolds, showing that both strain and curvature effects need to be considered when modeling the flame structure of turbulent premixed hydrogen-air flames prone to thermo-diffusive instabilities.

Future works should expand this analysis to various conditions, such as different equivalence ratios and higher turbulent intensities. 
Further, \textit{a-posteriori} analysis based on fully coupled simulations should be performed.
Additionally, when aiming for applications in the context of large eddy simulations, suitable sub-grid models for turbulent hydrogen combustion are of key importance. 

\section*{Acknowledgments} \addvspace{10pt}
The research leading to these results has received funding from the European Union’s Horizon 2020 research and innovation program under the Center of Excellence in Combustion (CoEC) project, grant agreement No 952181 and from the Federal Ministry of Education and Research (BMBF) and the state of Hesse as part of the NHR4CES Program.
The authors gratefully acknowledge the Gauss Centre for Supercomputing e.V. for funding this project by providing computing time on the GCS Supercomputer SuperMUC-NG at Leibniz Supercomputing Centre.

\newpage


\bibliography{publication.bbl}
\bibliographystyle{unsrtnat_mod}

\end{document}